# Beyond Accuracy: How Humans Evaluate Legally Correct but Socially Controversial Legal Advice from Machines

Benjamin Minhao Chen[a] Zhiyu Li[b1]

[a] Associate Professor, Faculty of Law, University of Hong Kong

[b] Associate Professor in Law and Policy, Durham Law School, Durham University

Abstract

AI systems are increasingly used to provide legal advice, raising questions about whether laypeople accept guidance from algorithms—especially when that advice is legally correct but socially controversial. We report a preregistered survey experiment with 3,348 adults in mainland China examining how people evaluate identical legal advice when it is attributed either to an AI system or to a human lawyer, and when it is accompanied by reasoning or not.

Contrary to expectations of algorithm aversion, attribution to an AI system has no net effect on perceived reasonableness. However, mediation analyses reveal opposing psychological pathways underlying this null result. AI-attributed advice is perceived as more objective, which increases perceived reasonableness, but also as less comprehensive and less attentive to special circumstances, which decreases perceived reasonableness. By contrast, providing legal reasoning substantially increases perceived reasonableness regardless of source, largely by enhancing perceptions of objectivity. Qualitative responses corroborate this tension between objectivity and contextual sensitivity in evaluations of legal advice.

Together, these findings suggest that public responses to AI legal advisors are shaped not by rigid attitudes toward automation, but by the balancing of competing normative expectations. The results have implications for theories of algorithm aversion and the design of AI recommendation systems in normatively salient domains.

Keywords: Algorithmic aversion, legal advice, attribution, reasoning, objectivity

Author Note: All authors have approved the submission and declare no conflicts of interest. The study protocol was approved by the relevant institutional ethics committees, and all participants provided informed consent before participation. This research was supported by the University of Hong Kong Lasting Impact Fund.

[1] Corresponding author. Durham University, Law School, PCL 176, The Palatine Centre, Stockton Rd., Durham, UK. E-mail address: zhiyu.li@durham.ac.uk.

## 1 INTRODUCTION

Artificial Intelligence (AI) systems are increasingly being designed to support and in some cases substitute for the opinion of trained experts. Among the various applications of generative AI, the dispensation of legal advice represents a use case that is important for the liberty and welfare of citizens. Everyone is assumed to know the law, and ignorance, as the saying goes, is no excuse. Yet the price of legal counsel, even for simple matters, can be exorbitant. Large Language Models (LLMs) and related technologies promise to make legal knowledge more accessible than ever before. By offering personalized information about rights, duties, and obligations, AI counsellors can assist individuals in planning their affairs and resolving their disputes. These benefits, however, can only be fully realized if ordinary people are amenable to AI-generated legal advice. Users may not act on legal information if they are not confident in the provider's neutrality or competence. They may also seek a second opinion, thereby incurring duplicative costs. An exploration of the factors that influence how laypeople process legal advice from AI systems is thus critical and timely.

We examine how ordinary people respond to AI systems as legal advisors. Although similar research questions have recently been broached by others (Hagan, 2024; Seabrooke, et al., 2024; Schneiders, et al., 2025), this study makes two distinct contributions to our understanding of human-computer interaction in the legal domain. First, it probes how individuals assess the reasonableness of advice ascribed to an AI system when the information, though correct from a legal perspective, is in tension with popular morality. Second, this study investigates the variables that mediate how individuals assess such advice when it is said to come from an AI system rather than a qualified lawyer. An experiment conducted on 3348 adults residing in the People's Republic of China (PRC) finds that an explanation enhanced perceptions of the reasonableness of legal advice regardless of its provenance. Overall, attribution had no impact on how the advice itself is evaluated. Even so, the evidence suggests differences in how subjects evaluated legal advice that was ascribed to a machine rather than a human. AI-attributed advice was more likely to be perceived as objective, thereby enhancing reasonableness judgments. However, it was also less likely to be seen as having taken all circumstances into account and more likely to neglect uniqueness, thereby eroding reasonableness judgments. These results imply a tension between objectivity on the one hand, and comprehensiveness of consideration and attention to uniqueness on the other.

## 2 ALGORITHMIC ADVICE

Originally invented to perform complex or tedious numerical operations, computers can now generate individualized recommendations for users in a variety of environments. How users react to these suggestions has been the subject of extensive research spanning several disciplines. The tendency to accept or incorporate machine advice is often referred to as "algorithm appreciation" (Logg, Minson, & Moore, 2019) or "automation bias" (Mosier, Skitka, Heers, & Burdick, 1998). The tendency to reject or ignore such recommendations, on the other hand, has been called "algorithm aversion" (Dietvorst, Simmons, & Massey, 2015) or "automation disuse" or "under-reliance" (Parasuraman & Riley, 1997). While these labels appear synonymous, they do not have identical connotations. Here, we use "algorithm aversion" to refer to a "biased assessment of an algorithm which manifests in negative behaviors and attitudes towards the algorithm compared to a human agent" (Jussupow, Benbasat, & Heinzl, 2020). We limit our attention to algorithm aversion in the context of advice, suggestions, or recommendations, rather than decisions.

The literature on algorithm aversion is characterized by heterogeneous and sometimes conflicting findings. One prominent account highlights the role of observed error. Indeed, the term algorithm aversion originally described a phenomenon where people react more negatively to mistakes committed by an algorithm than by a human (Jussupow, Benbasat, & Heinzl, 2020; Prahl & Van Swol, 2017). This penalty persists even when users witness algorithmic forecasts outperforming human ones, suggesting that it is machine imperfection rather than mere inaccuracy that produces aversion (Dietvorst, Simmons, & Massey, 2015). Further, aversion has been linked to the desire for control and agency. Individuals are more willing to rely on algorithmic recommendations when they are able to modify them, even minimally (Dietvorst, Simmons, & Massey, 2018; Bohlen, Kruschel, Rosenberger, Zschech, & Kraus, 2025).

More fundamentally, algorithm aversion may stem from visceral attitudes toward new technologies. Technological revolutions have sometimes been experienced as threats to established values and social order, thereby inducing societal-level anxiety or "moral panic" (Cohen, 1972; Goode & Ben-Yehuda, 1994; Critcher, 2008). AI systems might be viewed as unfamiliar and potentially dangerous agents whose behavior is "unknown" (Turel & Kalhan, 2023). This primal prejudice can be reduced through learning and experience (Turel & Kalhan, 2023). Overall, improving the perceived expertise of an automated assistant can turn algorithm aversion into appreciation (Hou & Jung, 2021; Daschner & Obermaier, 2022; de Jong, Bos, van Berkel, & Lamers, 2025).

Algorithmic aversion also depends on the nature of the task being performed. Algorithms are trusted and relied on less for tasks that are seen as subjective rather than objective, especially those seen as requiring emotional rather than purely cognitive capacities (Castelo, Bos, & Lehmann, 2019). This is partly because users draw on heuristics that portray machines as inflexible, impersonal, and unable to account for contextual details or special circumstances (Sundar, 2020). In healthcare, patients can be wary of automated diagnostics and treatment recommendations because of the absence of empathy (Riess & Kraft-Todd, 2014) or out of "a concern that AI providers are less able than human providers to account for their unique characteristics and circumstances" (Longoni, Bonezzi, & Morewedge, 2019). People seem more averse towards algorithmic advice in moral as opposed to non-moral matters (Arlinghaus, Straßmann, & Dix, 2025), although evidence for this claim is mixed (Liu & Wang, 2025). Additionally, algorithmic aversion has been observed in regard to creative pursuits. The same musical soundtrack or artwork is often rated as less creative when it is presented as being AI generated rather than human generated (Magni, Park, & Chao, 2024; Heimstad, Wien, & Gaustad, 2025). This form of "gatekeep[ing]" is driven by the perception that AI systems do not exert the effort and intentionality necessary for genuine creativity (Magni, Park, & Chao, 2024; Heimstad, Wien, & Gaustad, 2025).

Because aversion often reflects doubts about reliability, flexibility, and transparency, explanation is often proposed as a means of addressing user resistance toward automated advice. Explanations may increase trust in, acceptance of, or willingness to use an algorithmic recommendation, particularly when they clarify its basis or enhance beliefs about its quality (Kizilcec, 2016; Kunkel, Donkers, Michael, Barbu, & Ziegler, 2019; Ning, Lu, Li, & Gupta, 2024; Shin, 2021). However, greater intelligibility does not always overcome the concern that AI systems lack the human-like affective capacities required for subjective tasks, or the sensitivity to personal and situational nuances thought necessary for tailored guidance (Castelo, Bos, & Lehmann, 2019; Longoni, Bonezzi, & Morewedge, 2019).

## 3 THE LEGAL USE CASE

The emergence of Large Language Models (LLMs) has made automated legal advice more feasible and naturalistic than ever before. A qualitative study conducted in the United States reported significant public enthusiasm for using generative AI for legal assistance (Hagan, 2024), and survey evidence from China similarly suggests that residents are positively disposed toward AI-enabled legal services (Chen & Li, 2020). Perhaps this is not entirely surprising since neutrality and consistency—attributes that humans tend to associate with machines—are basic tenets of legality. There are, however, features of legal reasoning that could give rise to hesitation or reservations about AI advisors.

First, law is open textured. The notion of open texture is invoked by legal theorists to describe the inherent uncertainty of law (Waismann, 1945; Hart H. L., 1961). A legal rule contains a core of determinacy and a penumbra of indeterminacy (Hart H. L., 1961). For example, a rule banning vehicles from a park clearly precludes the entry of a fully operational passenger bus. Reasonable people can disagree, however, about whether the prohibition applies equally to baby tricycles or motorized wheelchairs. This indeterminacy is not a failure of drafting but an inevitable feature of language itself. Furthermore, unlike mathematical concepts that have exact definitions, legal terms like "reasonable," "due care," and "fair" do not have precise boundaries and are contested through their application to unforeseen circumstances in the light of competing values. The practice of law therefore demands sensitivity to the jurisprudential salience of novel facts and normative judgment. Interpreting and advising on law can involve subjectivity and creativity, making the task one that is liable to induce algorithm aversion.

Second, the law, as a system of hierarchical rules, can sometimes dictate results that contradict the moral values and principles of many citizens (Hart H. L., 1961). Although jurists continue to debate whether legal validity can ever depend on moral criteria, it is generally accepted that the legally mandated outcome in a situation does not have to mirror the morally mandated outcome (Hart H. L., 1958; Raz, 1979; Dworkin, 1986). This separation between law and morality raises the possibility of the same legal advice being evaluated more negatively if it is being rendered by an AI system rather than a human lawyer. A legal advisor that offers normatively questionable but professionally proper guidance may be perceived as being inaccurate, a belief that disproportionately impacts attitudes toward machines relative to humans.

These considerations invite empirical investigation into whether laypeople are in fact less receptive to legal advice attributed to AI systems than to human advisors. An experiment conducted on lawyers and law students in the Czech Republic revealed a preference for human-crafted documents over AI-generated ones (Harasta, Novotná, & Savelka, 2024). Another experiment, fielded in Kenya, reported that citizens considered legal decisions based on AI-produced research to be as legitimate as legal decisions based on human-conducted research (Flanagan, Almeida, Chen, & Gitahi, 2025). Perhaps the most germane to our present inquiry is a finding from the United Kingdom that ordinary people were more willing to act on machine-generated than human-generated legal advice (Seabrooke, et al., 2024). That experiment did not hold the text of legal advice constant and participants might have preferred LLMs to lawyers because their responses displayed higher levels of complexity and confidence. Moreover, none of these studies explore whether and how the tension between law and morality shapes lay evaluations of AI-generated legal advice.

This article presents a large *n* experiment examining how ordinary Chinese residents assess the reasonableness of advice in cases where the legally correct answer is socially controversial. Reasonableness refers to a normative judgment about whether a piece of legal advice is sensible, justified, and appropriate given the relevant facts and social norms of a case. It is conceptually distinct from trust—"a willingness to be vulnerable" (Mayer, Davis, & Schoorman, 1995), from credibility—the perceived expertness and trustworthiness of a source (Hovland, Janis, & Kelley, 1953), and from fairness—appropriate impartiality in the treatment of parties (Cullity, 2008). Unlike these source- or process-oriented constructs, reasonableness is a substantive evaluation of the advice itself rather than who gave it or how it was produced.

Our study manipulated attribution rather than the substance or style of legal advice, and the advice given was always accurate as a matter of law. Research on algorithm aversion suggests that ascribing advice to an AI system might depress reasonableness ratings, particularly where users doubt an algorithm's ability to account for context-specific factors. At the same time, because subjects were asked to evaluate the normative appeal of the advice rather than its legal soundness, they need not draw inferences from the source label at all. Formally, we test the following hypotheses.

*Hypothesis* 1: Subjects judge legal advice to be as reasonable when it is attributed to an AI system as when it is attributed to a human.

*Hypothesis* 2: Subjects judge legal advice to be more reasonable when it is accompanied by an explanation than when it is not.

*Hypothesis* 3: To the extent that participants judge legal advice to be less reasonable if it is attributed to an AI system rather than a human, this difference should be greater when the advice is not accompanied by an explanation.

To clarify the mechanisms that may underlie any effects of attribution and explanation, we also examine four candidate mediators. The first mediator, referred to as "mistake," captures perceptions of an advisor's competence and reliability. Prahl and Van Swol (2017) observe that people discount algorithmic forecasts more sharply than human forecasts after observing a single instance of poor performance, and Daschner and Obermaier (2022) conclude that beliefs about accuracy strongly predict trust in algorithmic recommendations. The second mediator, "objectivity," captures the perception of an advisor as impartial and free from subjective influences. Yalcin et al. (2022) find that machines are often seen as more objective and less emotionally biased than humans, and that this impression positively influences evaluations of algorithmic recommendations.

The third mediator, "attention to special circumstances," captures the perception that an advisor may have failed to recognize the features that make a case normatively different from otherwise like cases. Longoni, Bonezzi, and Morewedge (2019) demonstrate that resistance to AI guidance in medical contexts is driven by uniqueness neglect—the belief that an algorithm cannot adequately account for anomalous or atypical features of a case. Flanagan et al. (2025) similarly conjecture that doubts about acting on AI-generated legal analysis may stem from perceptions of AI decision-making as insufficiently responsive to contextual variation. The fourth mediator, "comprehensiveness," captures the perception that an advisor may have missed relevant elements of a case, even if the case is not idiosyncratic. It is conceptually related to but distinct from attention to special circumstances: where the latter concerns sensitivity to features that make a case normatively atypical, comprehensiveness concerns the thoroughness of factual processing in ordinary cases.

Together, these mediators relate to two broad psychological dimensions that prior research has shown to shape evaluations of algorithmic versus human advice: the perceived capability of the advisor and the degree of personalization the task demands (Qin, et al., 2025). Analysis of these potential mediators illuminates the normative expectations that subjects have of legal advice and how these expectations color their attitudes toward artificial lawyers. Attribution and reasoning may shape reasonableness judgments by changing participants' perceptions of the advice or advisor. Attribution to AI may increase perceived objectivity if participants associate machines with neutrality, but may also increase the perceived likelihood of mistake and reduce perceived comprehensiveness or attention to special circumstances if participants view AI systems as less sensitive to contextual nuance. Reasoning may reduce perceived mistake and increase perceived comprehensiveness, objectivity, and attention to special circumstances by making the legal basis for the advice more explicit. In turn, perceived mistake should make advice appear less reasonable, whereas perceived comprehensiveness, objectivity, and attention to special circumstances should make advice appear more reasonable.

## 4 SCENARIO SELECTION SURVEYS

We started by conducting surveys in Mandarin Chinese to measure the controversiality of legally mandated outcomes in several legal cases. The purpose of these surveys is to select suitably controversial scenarios for the main experiment testing the effect of attribution on lay evaluations of the reasonableness of legal advice.

## 4.1 Method

### *4.1.1 Design*

In the first survey, each participant encountered four legal cases: one about an offending motorist's refusal to sign a traffic accident report, one about the return of a betrothal gift after a short marriage, one about an agreement between sons about filial support, and one about the death of a non-paying passenger in an accident. In the second survey, each participant encountered two legal cases: one about the right of a child born outside marriage to an equal division of an estate and one about tortfeasor's liability for a victim's pre-existing medical condition.

For each of these items, participants read a stylized account of the facts. Participants were then asked which one of two binary outcomes were, in their opinion, more reasonable. They were also asked to indicate their levels of confidence in their judgments on a 5-point scale. Participants did not receive any information about legal rules or principles. Cases were presented in the same order to all participants.

The facts of all six cases are reproduced in Figure 1. For example, in the first case of the first survey, participants read about a traffic accident between two motorists. Briefly, on February 1, 2023, Yu caused a traffic accident by running a red light and colliding with Xiao's van. He was found to be fully at fault. The official repair cost was set at 12,300 yuan. However, Yu refused to sign the accident and damage reports, and subsequent negotiations to resolve the matter were unsuccessful. Participants were asked whether they thought it was more reasonable for Xiao to be awarded or not awarded 12,300 yuan in compensation despite the absence of He's signature on the reports.

| Case Facts | Binary Outcomes |
|---|---|
| **Case 1:** On the evening of February 1, 2023, Yu, on his way home from work, violated traffic rules by running a red light, causing his small sedan to collide with Xiao's van. After determination by the Traffic Police Brigade's accident responsibility assessment, Yu should bear full responsibility for this accident. Through joint damage assessment by the Traffic Police Brigade, insurance company, and auto repair shop, it was determined that the van's repair costs resulting from this accident were 12,300 yuan, and they provided Yu and Xiao with the traffic accident determination report and damage assessment report. Afterwards, Yu refused to sign the traffic accident determination report and damage assessment report, and after multiple negotiations, no agreement was reached. | Which of the following results do you think is more reasonable?<br>1. Xiao obtained compensation of 12,300 yuan for vehicle repair costs.*<br>2. Xiao was unable to obtain compensation of 12,300 yuan for vehicle repair costs. |
| **Case 2:** In January 2020, Xu and Zhu met through introduction and established a romantic relationship. One month later, both parties handled marriage registration procedures during their return to their hometowns for the New Year. Although Xu's family had difficult living conditions, they still gave Zhu a bride price of 50,000 yuan before marriage based on custom. After marriage, because both parties worked in other places, they never lived together and often quarreled endlessly over trivial matters on the phone. In February 2023, both parties handled divorce procedures at the Civil Affairs Bureau due to emotional breakdown. However, both parties disputed endlessly over the issue of returning the bride price. The male party Xu believed that the bride price was given according to custom, but both parties had no emotional foundation, never lived together after marriage, and had already handled divorce procedures, so Zhu should return the bride price money. Zhu believed that the bride price was Xu's voluntary gift and should not be returned. | Which of the following results do you think is more reasonable?<br>1. Zhu returns the bride price money paid by Xu.*<br>2. Zhu does not return the bride price money paid by Xu. |
| **Case 3:** Chen (currently 70 years old) divorced her ex-husband Guo in September 2016. They had two sons together, Guo Da (currently 40 years old) and Guo Er (currently 30 years old). During the divorce period, the eldest son Guo Da and the second son Guo Er signed a support agreement without the consent of their mother Chen, stipulating: "The eldest son Guo Da supports mother Chen, and the second son Guo Er supports father Guo." After retirement, mother Chen had no source of income, lived alone continuously and suffered from multiple diseases, requiring high medical expenses. In September 2023, Chen contacted her second son Guo Er, requesting that Guo Er bear her medical expenses. The second son Guo Er refused Chen's request, stating that the support agreement had already stipulated that mother Chen would be supported by the eldest son Guo Da, and the father would be supported by himself. He had already supported his father according to the agreement and could not accept taking responsibility for supporting his mother together with the eldest son Guo Da. | Which of the following results do you think is more reasonable?<br>1. Mother Chen obtains medical expenses paid by second son Guo Er.*<br>2. Mother Chen cannot obtain medical expenses paid by second son Guo Er. |
| **Case 4:** On January 6, 2022, after attending his friend's wedding, Yu drove his non-commercial small car home. Based on goodwill, Yu agreed to give his neighbor Zhong, who also attended the wedding, a free ride home. During the driving process, a traffic accident occurred, causing the death of passenger Zhong. Later, the traffic police department determined that although Yu had no intention or gross negligence, he bore full responsibility for the accident. | Which of the following results do you think is more reasonable?<br>1. Reduce the compensation liability of traffic accident perpetrator Yu for the losses arising from the death of a non-paying passenger, Zhong.*<br>2. Do not reduce the compensation liability of traffic accident perpetrator Yu for the losses arising from the death of a non-paying passenger, Zhong. |
| `**Case 5:** In June 2023, Xu died in an accident. During his lifetime, Xu had been married to his wife Li for thirty years and had a daughter. Many years ago, Xu met a single woman at a bar and had a one-night stand. The woman later became pregnant with Xu's child and, without Xu's knowledge, gave birth to a male infant Wang in May 2002. One week after Xu's death, Wang, then twenty-one years old, approached Li and Li's daughter to assert his inheritance rights to Xu's estate. Xu had not left any will during his lifetime. The parties disputed endlessly in multiple negotiations and failed to reach a unanimous agreement. | Which of the following results do you think is more reasonable?<br>1. Wang obtains an equal share of the inheritance as Xu's daughter.*<br>2. Wang cannot obtain an equal share of the inheritance as Xu's daughter. |
| **Case 6:** In March 2023, Wang was speeding on a highway and rear-ended the vehicle of Lü ahead. Lü suffered from congenital heart disease, and the occurrence of the rear-end accident caused Lü to be severely frightened and die on the spot. The accident was determined by the local traffic police brigade: Wang bore full responsibility for the accident, and Lü bore no responsibility. Lü's family members demanded that Wang bear all tort liability for the death consequences of Lü caused by his rear-end accident, totaling 500,000 yuan. | Which of the following results do you think is more reasonable?<br>1. Lü's own congenital heart disease can reduce Wang's tort liability for the rear-end accident.<br>2. Lü's own congenital heart disease cannot reduce Wang's tort liability for the rear-end accident.* |

Figure 1: Facts of six legal cases received by participants and the binary outcomes from which the participants were asked to choose. Outcomes are marked with an asterisk (*) represents the legally mandated outcomes of the legal cases.

*4.1.2 Participants*

The first survey was administered online to 210 adults residing in the PRC mainland between 4 and 14 April 2024. Participants were recruited by a professional survey company, Wei Diao Cha, and quota sampled to approximate the Chinese internet population for age and gender. 106 participants identified as female and 104 as male. Omitting two entries containing typographical errors in birth dates or extreme values, 19.7% of participants reported themselves as being in the 20 to 29 age range, 24.5% in the 30 to 39 age range, 20.2% in the 40 to 49 age range, 23.6% in the 50 to 59 age range, and 12.0% in the 60 to 79 age range. 64.8% of participants reported living in an urban area and 35.2% in a rural area.

The second survey was administered online to 201 adults residing in the PRC mainland between 29 May and 11 June 2024. Participants were recruited by Wei Diao Cha and quota sampled to approximate the Chinese internet population for age and gender. Participants in the first survey were excluded from taking part in the second survey. 95 participants identified as female and 106 as male. Omitting two entries containing typographical errors in birth dates, 16.1% of participants reported themselves as being in the 20 to 29 age range, 23.6% in the 30 to 39 age range, 20.6% in the 40 to 49 age range, 24.1% in the 50 to 59 age range, and 15.6% in the 60 to 79 age range. 70.1% of participants reported living in an urban area and 29.9% in a rural area.

### 4.2 Results

The results of the two surveys are summarized in Figure 2. For the first survey, (1) the traffic accident report case split respondents 84.3% to 15.7%, the majority believing an award of damages to the innocent motorist be more reasonable; (2) the betrothal gift case split respondents 66.7% to 33.3%, the majority believing the return of the gift to the estranged husband to be more reasonable; (3) the filial support case split respondents 79.0% to 21.0%, the majority believing the mother's entitlement to contribution from her younger son to be more reasonable; and (4) the pre-existing condition case split respondents 77.1% to 22.9%, the majority believing the reduction of liability for the driver to be more reasonable. For the second survey, (5) the estate division case split respondents 60.7% to 39.3%, the majority believing a non-equal share for the child born outside marriage to be more reasonable; and (6) the pre-existing medical condition case split respondents 60.2% to 39.8%, the majority believing the non-reduction of liability for the driver to be more reasonable.

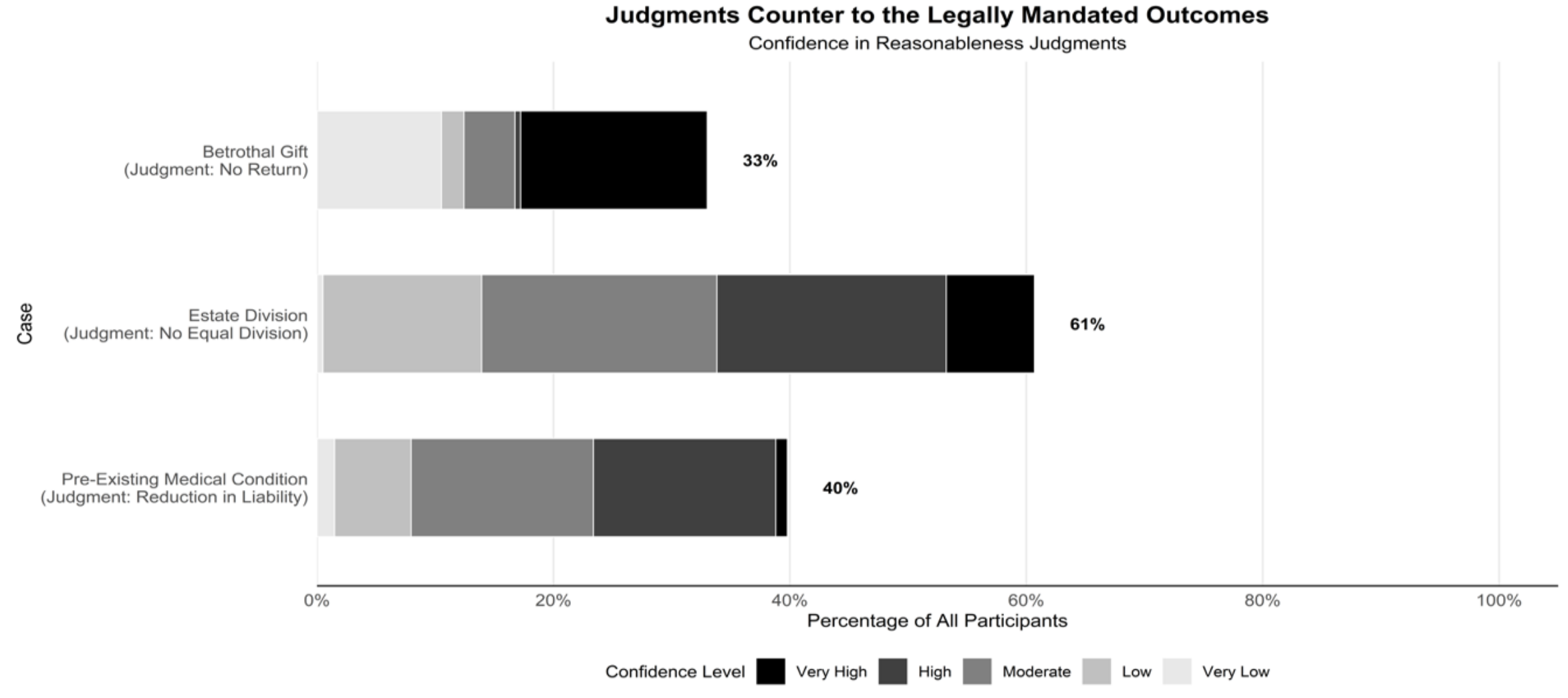

Figure 2: Distribution of participant confidence when making reasonableness judgments that contradict legal requirements across three case scenarios.

Of the six cases tested, three produced approximately 60-40 splits in opinion: the betrothal gift case, the estate division case and the pre-existing medical condition case. In the estate division case, a majority of participants judged what happened to be the legally mandated outcome on the facts to be less reasonable than the opposing alternative. In the betrothal gift case and pre-existing medical condition case, a majority of participants judged what happened to be the legally mandated outcomes on the facts to be more reasonable than the opposing alternatives.

## 5 MAIN EXPERIMENT

The main experiment administered in Mandarin Chinese tests whether ordinary people rate the reasonableness of accurate but socially controversial legal advice differently if it is said to be generated by an AI system rather than a human lawyer. Hypotheses, participant exclusion criteria, and statistical analyses were preregistered on the Open Science Framework.

### 5.1 Method

#### *5.1.1 Data Availability Statement*

Deidentified data, survey materials, and analysis code will be made available on OSF. The OSF preregistration is available at https://osf.io/gx9pa/overview.

#### *5.1.2 Ethical Approval and AI Use Statement*

This study was approved by the Ethical Review Boards of the University of Hong Kong (EA230601) and Durham University (LAW-2024-06-30T16_33_13-kszf79). All participants provided informed consent prior to participation and were free to withdraw at any time. We used Codex and Claude.ai to replicate all analyses reported in this article and to proofread the final manuscript.

#### *5.1.3 Design*

The main experiment features a 2×2 factorial between-subjects design. The first factor is attribution of the legal advice: AI system or human lawyer. The second factor is the reasoning given in the legal advice: reasoning or no reasoning. Participants are randomly sorted to one of three scenarios: the betrothal gift scenario, the estate division scenario, and the pre-existing medical condition scenario. For each scenario, participants read the facts of the case and legal advice ascribed to either an AI system or a human lawyer. The legal advice may contain reasoning based on relevant laws and regulations or it may contain no reasoning at all. The legal advice presented to participants is constant across the attribution factor: specifically, participants assigned to the AI system, no reasoning condition and participants assigned to the human lawyer, no reasoning condition received identical legal advice. Participants assigned to the AI system, reasoning condition and participants assigned to the human lawyer, reasoning condition also received identical legal advice. In both the reasoning and no reasoning conditions, the legal advice was accurate, that is, the conclusion reached was correct under existing law. For example, participants assigned to the estate division case were informed that the opinion of Lawyer Li or the AI legal service platform was that the child born outside marriage was entitled to an equal share of the inheritance. Figure 3 shows variations of the vignette for the estate division case.

Vignette (Human Attribution): In June 2023, Xu died in an accident. During his lifetime, Xu had been married to his wife Li for thirty years and had a daughter. Years ago, Xu met a single woman at a bar and had a one-night stand. The woman subsequently became pregnant with Xu's child and, without Xu's knowledge, gave birth to a male infant Wang in May 2002. One week after Xu's death, Wang, then twenty-one years old, approached Li and Li's daughter to claim his inheritance rights to Xu's estate. Xu had not left any will during his lifetime. The parties engaged in multiple negotiations but were in dispute and could not reach a unanimous agreement.

Wang then sought legal help from [*a local law firm*]. [*Lawyer Li*] analyzed the dispute facts, relevant evidence, and laws and regulations, and believed that Wang had a good chance of obtaining an equal share of the inheritance as Xu's daughter through litigation.

Vignette (AI Attribution): In June 2023, Xu died in an accident. During his lifetime, Xu had been married to his wife Li for thirty years and had a daughter. Years ago, Xu met a single woman at a bar and had a one-night stand. The woman subsequently became pregnant with Xu's child and, without Xu's knowledge, gave birth to a male infant Wang in May 2002. One week after Xu's death, Wang, then twenty-one years old, approached Li and Li's daughter to claim his inheritance rights to Xu's estate. Xu had not left any will during his lifetime. The parties engaged in multiple negotiations but were in dispute and could not reach a unanimous agreement.

Wang then sought legal help from [*an online intelligent service platform*]. [*The AI legal service platform*] analyzed the dispute facts, relevant evidence, and laws and regulations, and believed that Wang had a good chance of obtaining an equal share of the inheritance as Xu's daughter through litigation.

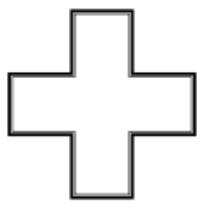

Reasoning: according to the provisions of Article 1127 of the Civil Code, "inheritance shall be in the following order: (1) First order: spouse, children, parents... The term 'children' as used in this Part includes legitimate children, illegitimate children, adopted children, and stepchildren with a dependency relationship," and Article 1071 of the Civil Code, which provides that "illegitimate children enjoy the same rights as legitimate children, and no organization or individual may harm or discriminate against them."

Figure 3: Vignettes with Human and AI Attributors and Reasoning Conditions of the Estate Division Case

After reading the scenario, participants were posed two comprehension check questions. The comprehension check questions related to the facts of the case and not the legal advice given. Participants were then asked to judge the reasonableness of the legal advice given on a 1 to 7 scale and to provide a short justification for their answers. The reasonableness rating is the outcome measure of interest. They were also asked to indicate, on a 1 to 7 scale, (1) how likely they thought the human lawyer or AI system would make analytical errors when providing legal advice, (2) how likely they thought the human lawyer or AI system would take all the factual elements in the case into consideration when providing legal advice, (3) how likely they thought the legal advice provided by the human lawyer or AI system was objective, and (4) to what extent they thought the human lawyer or AI system took into account special circumstances that may exist in the case when providing legal advice. These measures are candidate mediators of the outcome measure.

Finally, participants were administered a short multiple-choice quiz comprising six questions about basic legal facts. This quiz measures participants' level of legal knowledge. Demographic information on age, gender, education, and other data relating to experience in using AI legal software and perceptions of the reliability of AI legal software were also collected.

*5.1.4 Participants*

The main experiment survey was administered online to 5,040 adults residing in the PRC mainland between 10 August and 12 December 2024. Participants were recruited by Wei Diao Cha and quota sampled to approximate the Chinese internet population for age and gender. Excluding participants who failed the comprehension checks leaves a sample of 3,348 participants. Of these participants, 1712 (51.1%) identified as male, 1633 (48.8%) as female, and 3 (0.1%) as other. Omitting 17 entries containing non-numeric birth dates and 24 entries containing impossible or extreme age values, 32.0% of participants reported themselves as being in the 20 to 29 age range, 33.7% in the 30 to 39 age range, 21.5% in the 40 to 49 age range, 9.5% in the 50 to 59 age range, and 3.3% in the 60 to 79 age range. 39.9% of participants stated that they had education at or below the level of a three-year undergraduate degree, 52.8% stated they had education at the level of a four-year undergraduate degree, 6.8% stated they had education at the level of a master's degree, and 0.5% stated they had education at the level of a doctoral degree. 7.2% of participants claimed a major in law or legal studies; the remainder did not.

Descriptive statistics and pairwise correlations for quantitative variables are reported in Table 1.

Table 1. Descriptive Statistics and Correlations for Quantitative Variables

| Variable | N | M | SD | 1 | 2 | 3 | 4 | 5 | 6 | 7 |
|---|---|---|---|---|---|---|---|---|---|---|
| 1. Age | 3,307 | 36.27 | 11.07 | — | | | | | | |
| 2. Legal knowledge | 3,348 | 4.65 | 1.19 | 0.03 | — | | | | | |
| 3. Reasonableness | 3,348 | 5.11 | 1.38 | 0.07 | 0.03 | — | | | | |
| 4. Mistake | 3,348 | 3.32 | 1.52 | 0.04 | -0.12 | -0.17 | — | | | |
| 5. Comprehensiveness | 3,348 | 4.81 | 1.34 | 0.08 | -0.01 | 0.44 | -0.10 | — | | |
| 6. Objectivity | 3,348 | 5.06 | 1.26 | 0.08 | 0.10 | 0.55 | -0.12 | 0.47 | — | |
| 7. Attention to Special Circumstances | 3,348 | 4.31 | 1.51 | 0.10 | -0.05 | 0.38 | 0.02 | 0.52 | 0.35 | — |

Note. Values below the diagonal are Pearson correlations computed using pairwise complete observations. Age descriptives and age correlations use valid ages only: missing, non-numeric, impossible, or extreme values have been excluded.

## 5.2 Data Analysis

*5.2.1 Main Effect of Attribution*

To analyze the effect of attribution on reasonableness judgments in each of the scenarios, we estimate a linear regression model that has reasonableness ratings as the dependent variable and an indicator variable for an AI system attribution as the independent variable and compute HC2 robust standard errors. The estimated coefficient for the indicator variable can be interpreted as the average treatment effect (ATE) in a Neyman-Rubins framework (Imbens & Rubin, 2015). The accompanying standard errors are equivalent to the conservative Neyman standard errors in this framework (Imbens & Rubin, 2015). Cohen's *d* was computed by dividing each average treatment effect by the pooled standard deviation of reasonableness ratings in the corresponding two-group comparison.

For the betrothal gift scenario, the average reasonableness rating for participants in the human lawyer condition is 5.166 (SE = 0.056). Attribution of legal advice to an AI system has an ATE of -0.064 (SE = 0.074, p = 0.389, d = -0.05). For the estate division scenario, the average reasonableness rating for participants in the human lawyer condition is 4.644 (SE = 0.074). Attribution of legal advice to an AI system has an ATE of 0.044 (SE = 0.103, p = 0.665, d = 0.03). For the pre-existing medical condition scenario, the average reasonableness rating for participants in the human lawyer condition is 5.368 (SE = 0.052). Attribution of legal advice to an AI system has an ATE of 0.039 (SE = 0.072, p = 0.589, d = 0.03). Pooling across all scenarios, the average reasonableness rating for participants in the human lawyer condition is 5.104 (SE

= 0.035). Attribution of legal advice to an AI system has an ATE of 0.012 (SE = 0.048, p = 0.801, d = 0.01). Figure 4 presents the effect of attribution condition on reasonableness ratings across the three legal scenarios.

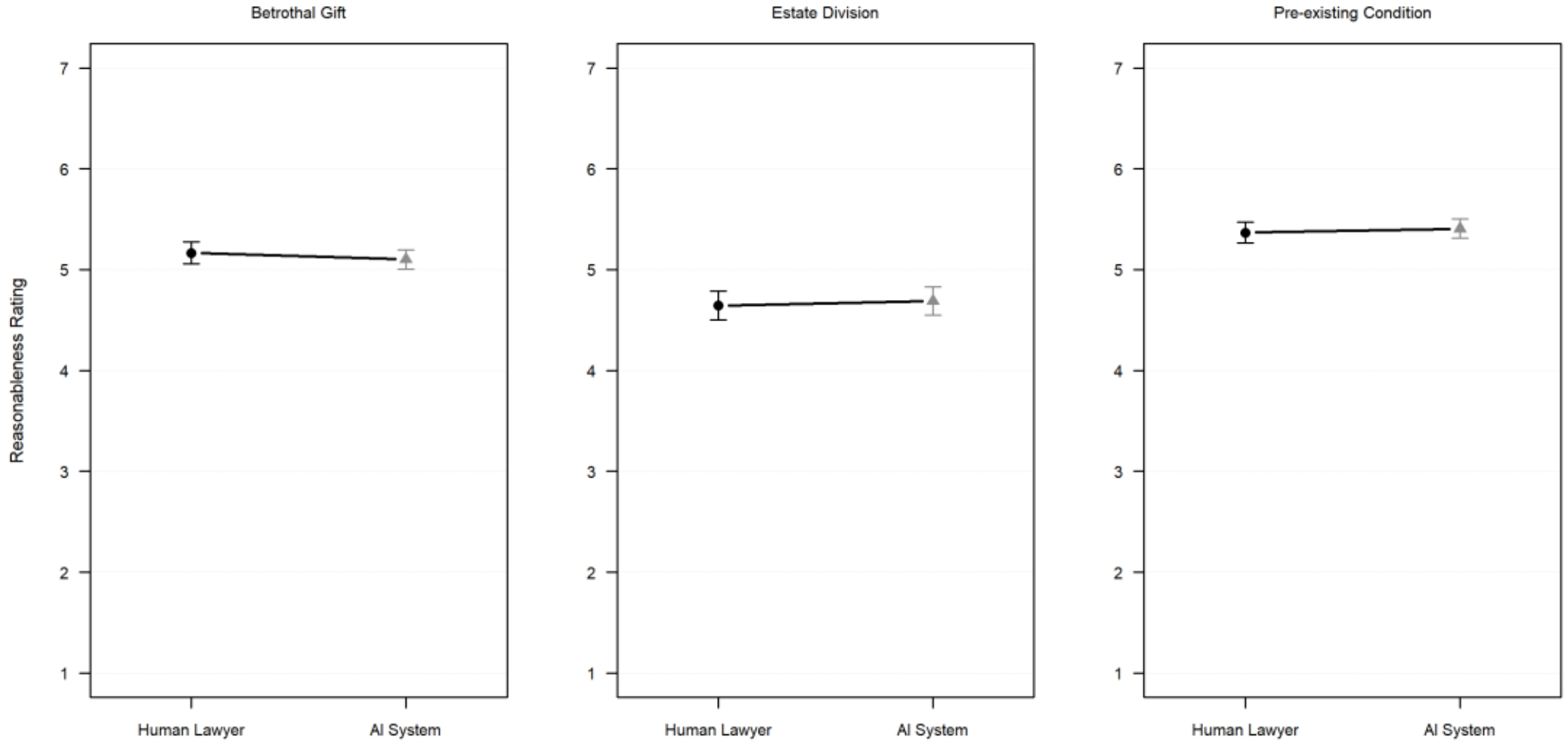

Figure 4: Effect of attribution condition on reasonableness ratings by legal scenario. Source attribution had no detectable effect on average reasonableness ratings for all three scenarios.

Overall, there is no evidence that attribution of legal advice to an AI system rather than a human lawyer had an effect on reasonableness ratings.

*5.2.2 Main Effect of Reasoning*

To analyze the effect of reasoning on reasonableness judgments in each of the scenarios, we estimate a linear regression model that has reasonableness ratings as the dependent variable and an indicator variable for reasoning as the independent variable and compute HC2 robust standard errors. For the betrothal gift scenario, the average reasonableness rating for participants in the no reasoning condition is 4.955 (SE = 0.055). Addition of reasoning has an ATE of 0.347 (SE = 0.073, p = 0.000, d = 0.29). For the estate division scenario, the average reasonableness rating for participants in the no reasoning condition is 4.416 (SE = 0.073). Addition of reasoning has an ATE of 0.519 (SE = 0.101, p = 0.000, d = 0.34). For the pre-existing medical condition scenario, the average reasonableness rating for participants in the no reasoning condition is 5.371 (SE = 0.048). Addition of reasoning has an ATE of 0.034 (SE = 0.072, p = 0.637, d = 0.03). Pooling across all scenarios, the average reasonableness rating for participants in the no reasoning condition is 4.974 (SE = 0.034). Addition of reasoning has an ATE of 0.274 (SE = 0.048, p = 0.000, d = 0.20). Figure 5 presents the effect of the reasoning condition on reasonableness ratings across the three scenarios.

Overall, there is evidence that addition of reasoning had a positive effect on reasonableness ratings.

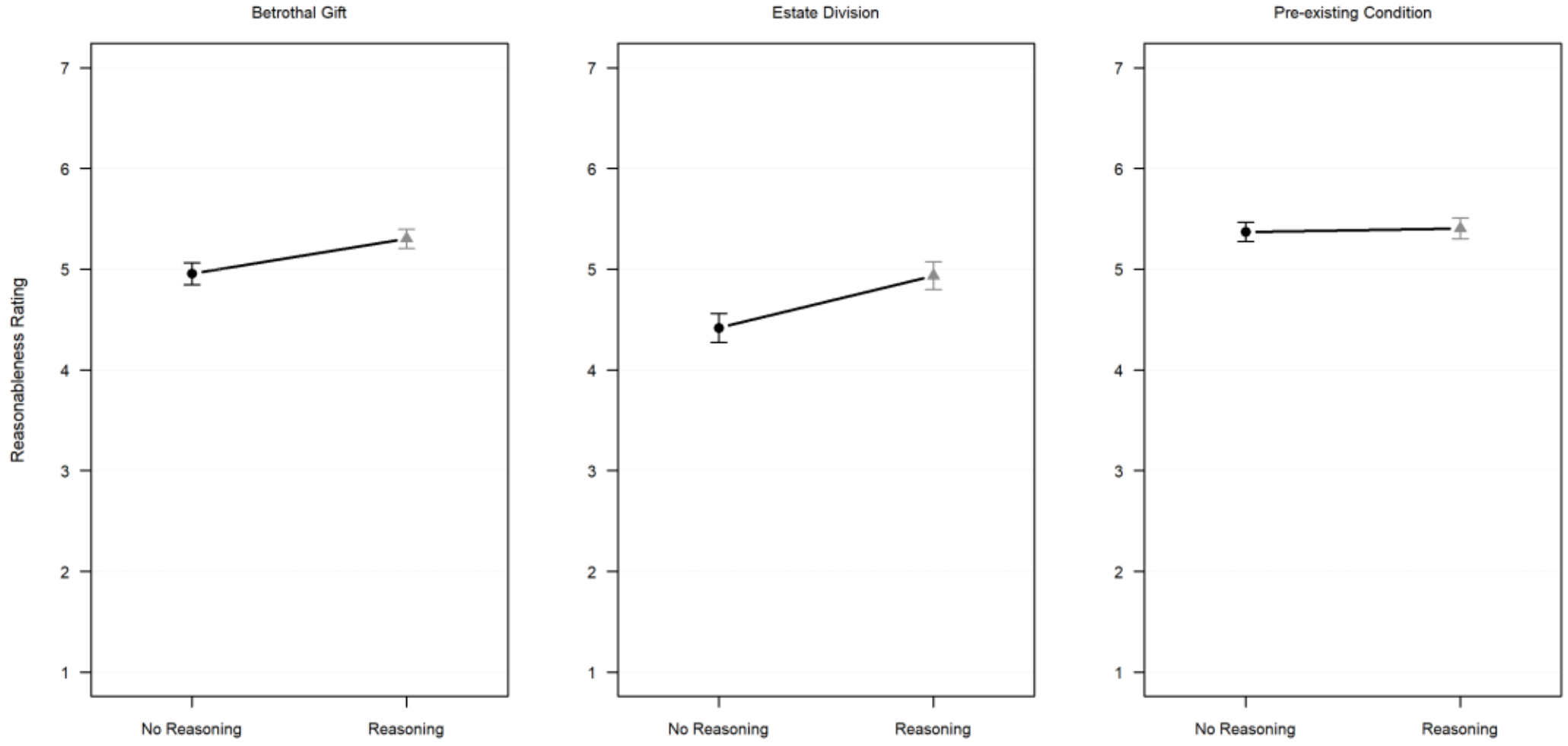

Figure 5: Effect of reasoning condition on reasonableness ratings by legal scenario. Providing reasoning significantly enhanced average reasonableness ratings for two of three scenarios, the largest effect being observed in the estate division case.

*5.2.3 Interaction Effect Between Attribution and Reasoning*

To analyze the interaction between attribution and reasoning on reasonableness judgments in each of the scenarios, we estimate a linear regression model that has reasonableness ratings as the dependent variable. The independent variables are an indicator variable for an AI system attribution, an indicator variable for reasoning, and an interaction term between the two indicator variables. HC2 robust standard errors are computed. The estimated coefficient for the indicator variables can be interpreted as the causal average treatment effect (CATEs) in a Neyman-Rubins causal framework (Lu, 2016).

For the betrothal gift scenario, the average reasonableness rating for participants in the human lawyer, no reasoning condition is 4.933 (SE = 0.084). Compared to this reference group, attribution of legal advice to an AI system has an ATE of 0.043 (SE = 0.111, p = 0.700) and addition of reasoning has an ATE of 0.458 (SE = 0.110, p = 0.000). Attribution of legal advice to an AI system and addition of reasoning has an ATE of -0.209 (SE = 0.147, p = 0.155) over and above the main effects.

For the estate division scenario, the average reasonableness rating for participants in the human lawyer, no reasoning condition is 4.388 (SE = 0.108). Compared to this reference group, attribution of legal advice to an AI system has an ATE of 0.057 (SE = 0.147, p = 0.700) and addition of reasoning has an ATE of 0.523 (SE = 0.146, p = 0.000). Attribution of legal advice to an AI system and addition of reasoning has an ATE of -0.005 (SE = 0.202, p = 0.979) over and above the main effects.

For the pre-existing medical condition scenario, the average reasonableness rating for participants in the human lawyer, no reasoning condition is 5.393 (SE = 0.069). Compared to this reference group, attribution of legal advice to an AI system has an ATE of -0.049 (SE = 0.096, p = 0.605) and addition of reasoning has an ATE of -0.056 (SE = 0.105, p = 0.596). Attribution of legal advice to an AI system and addition of reasoning has an ATE of 0.172 (SE = 0.144, p = 0.230) over and above the main effects.

Pooling across all scenarios, the average reasonableness rating for participants in the human lawyer, no reasoning condition is 4.982 (SE = 0.051). Compared to this reference group, attribution of legal advice to an AI system has an ATE of -0.017 (SE = 0.069, p = 0.804) and addition of reasoning has an ATE of 0.252 (SE = 0.070, p = 0.000). Attribution of legal advice to an AI system and addition of reasoning has an ATE of 0.043 (SE = 0.095, p = 0.650) over and above the main effects.

Overall, there is no evidence of an interaction effect between attribution of legal advice to an AI system rather than a human lawyer and addition of reasoning. Figure 6 presents the interaction between attribution and reasoning conditions across the three legal scenarios.

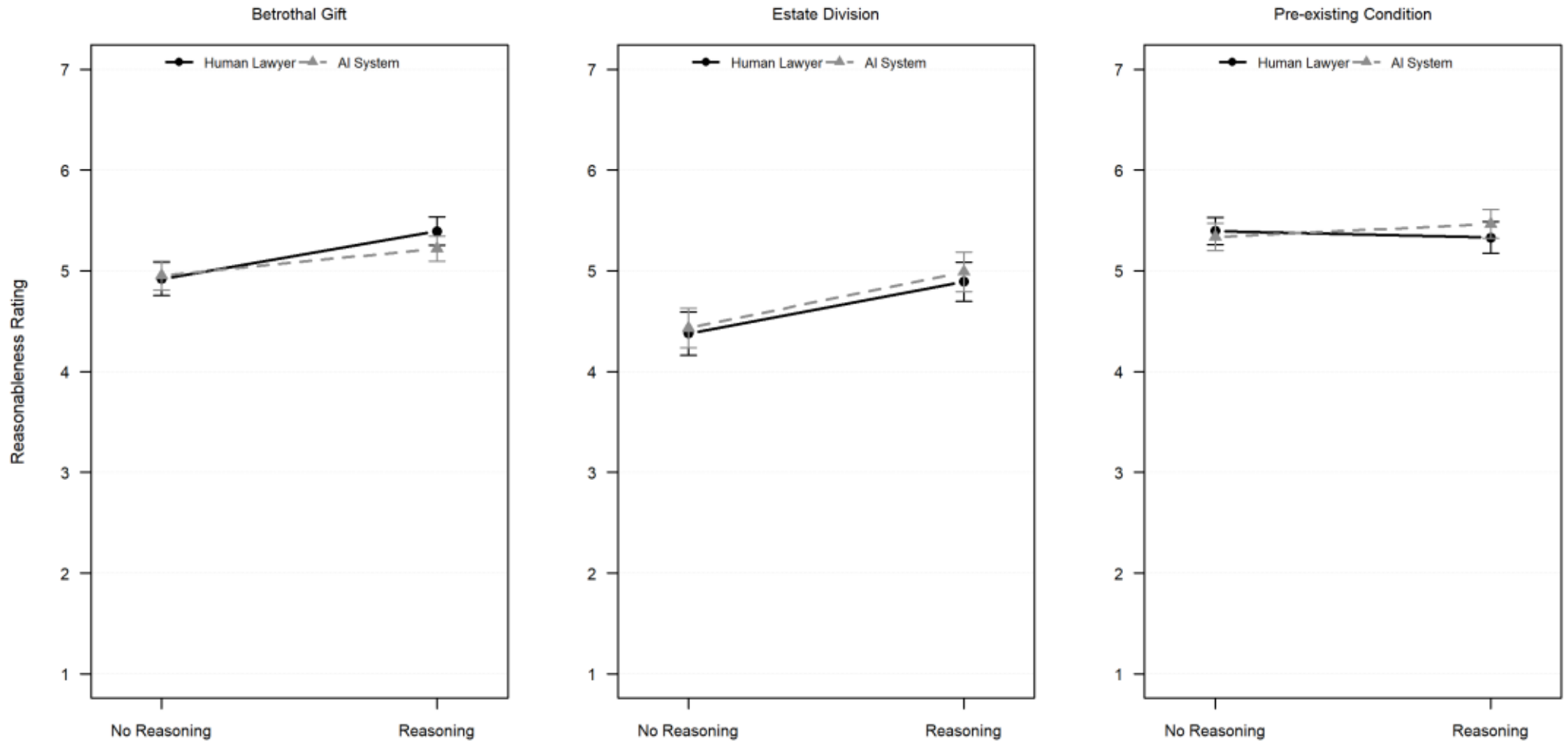

Figure 6: Interaction between attribution and reasoning conditions on reasonableness ratings by legal scenario. There is no significant interaction between source attribution and the provision of reasoning in any of the three scenarios.

### *5.2.4 Moderation by Legal Knowledge*

The legal knowledge quiz consisted of six questions. 9 or 0.3% of participants received a score of zero on the quiz, 1.3% received a score of one, 3.6% received a score of two, 10.6% received a score of three, 24.9% received a score of four, 31.6% received a score of five, and 27.9% received a score of six. The median score is five. As pre-registered, participants are coded as having high legal knowledge if they scored above the median and as having low legal knowledge if they scored at or below the median. To analyze heterogeneity in treatment effects by legal knowledge, we re-estimate the linear regression models above, introducing an additional indicator variable for high legal knowledge and interaction term or terms between this indicator variable and the other dependent variables. Regression results are presented in Table 2.

| *Variables* | Model 1 | Model 2 | Model 3 |
|---|---|---|---|
| ***(Intercept)*** | 5.073*** (0.043) | 4.947*** (0.041) | 4.969*** (0.061) |
| ***Agent (AI = 1)*** | 0.012 (0.058) | — | -0.044 (0.082) |
| ***Reasoning (Present = 1)*** | — | 0.271*** (0.057) | 0.221** (0.085) |
| ***Legal Knowledge (High = 1)*** | 0.108 (0.074) | 0.100 (0.074) | 0.048 (0.108) |
| ***Agent × Legal Knowledge*** | 0.005 (0.103) | — | 0.104 (0.148) |
| ***Reasoning × Legal Knowledge*** | — | 0.002 (0.102) | 0.096 (0.148) |
| ***Agent × Reasoning*** | — | — | 0.097 (0.115) |
| ***Agent × Reasoning × Legal Knowledge*** | — | — | -0.185 (0.204) |
| ***N*** | 3,348 | 3,348 | 3,348 |
| ***R² / Adjusted R²*** | 0.001 / 0.000 | 0.011 / 0.010 | 0.011 / 0.009 |
| ***F-statistic*** | 1.566 | 12.37*** | 5.537*** |

* $p < 0.05$, ** $p < 0.01$, *** $p < 0.001$

Table 2. Treatment Effects on Reasonableness Ratings by Legal Knowledge

Overall, there is no evidence that legal knowledge moderates the main effect of attribution on reasonableness ratings, the main effect of reasoning on reasonableness ratings, or the interaction effect between attribution and reasoning.

*5.2.5 Mediation Analysis*

Mediation analyses were conducted using a path-analytic approach following (Hayes, 2013), estimated via structural equation modelling (Rosseel, 2012).[2] Each model included all four candidate mediators simultaneously, with covariate adjustments for age, gender, level of education, study of law-related fields, and legal knowledge. Point estimates and 95% bias-corrected bootstrap confidence intervals are based on 5,000 bootstrap samples. Indirect effects whose confidence intervals exclude zero are considered statistically significant. The causal interpretation of mediation analysis relies on the assumption of no unmeasured confounding of the mediator-outcome relationship, which cannot be fully guaranteed in the absence of experimental manipulation of the mediators (Spencer, Zanna, & Fong, 2005; VanderWeele, 2015).

For the mediation analyses, participants with missing or non-numeric birth years and participants with impossible or extreme age values were excluded before estimating the models, leaving 3,307 observations.

The results of the mediation analysis for the effect of attribution on reasonableness ratings are presented in Table 3. The total effect in the covariate-adjusted mediation model is 0.009. The indirect effect through mistake is -0.018 with a

[2] Note that the preregistration specified that mediation analysis would be conducted following the framework of Imai, Keele, and Yamamoto (2010). The present approach, taken at the request of the journal, estimates all mediators simultaneously in a single structural equation model. The results are qualitatively similar across both methods.

95% CI of [-0.032, -0.007]. The indirect effect through comprehensiveness is -0.057 with a 95% CI of [-0.080, -0.038]. The indirect effect through objectivity is 0.049 with a 95% CI of [0.011, 0.088]. The indirect effect through attention to special circumstances is -0.061 with a 95% CI of [-0.085, -0.043]. Attribution to an AI system therefore decreased reasonableness ratings through diminished perceptions of attention to special circumstances and comprehensiveness, and enhanced perceptions of mistake. These negative indirect effects were partially offset by an enhanced perception of objectivity, which increased reasonableness ratings.

| | Estimate | 95% CI Lower Bound | 95% CI Higher Bound |
|---|---|---|---|
| Mistake | -0.018 | -0.032 | -0.007 |
| Comprehensiveness | -0.057 | -0.080 | -0.038 |
| Objectivity | 0.049 | 0.011 | 0.088 |
| Attention to Special Circumstances | -0.061 | -0.085 | -0.043 |

Table 3. Estimated Indirect Effects for the Effect of Attribution on Reasonableness Ratings

The results of the mediation analysis for the effect of reasoning on reasonableness ratings are presented in Table 4. The total effect in the covariate-adjusted mediation model is 0.284. The indirect effect through mistake is 0.014 with a 95% CI of [0.003, 0.027]. The indirect effect through comprehensiveness is 0.019 with a 95% CI of [0.006, 0.036]. The indirect effect through objectivity is 0.085 with a 95% CI of [0.047, 0.124]. The indirect effect through attention to special circumstances is -0.004 with a 95% CI of [-0.019, 0.011]. It seems therefore that 30.0% of the total effect of reasoning on reasonableness ratings can be explained by enhanced perceptions of objectivity, 6.8% by enhanced perceptions of comprehensiveness, and 5.0% by diminished perceptions of mistake.

| | Estimate | 95% CI Lower Bound | 95% CI Higher Bound |
|---|---|---|---|
| Mistake | 0.014 | 0.003 | 0.027 |
| Comprehensiveness | 0.019 | 0.006 | 0.036 |
| Objectivity | 0.085 | 0.047 | 0.124 |
| Attention to Special Circumstances | -0.004 | -0.019 | 0.011 |

Table 4. Estimated Indirect Effects for the Effect of Reasoning on Reasonableness Ratings

The component paths clarify how the indirect effects arise. In the attribution model, AI attribution increased perceived mistake and objectivity, but decreased perceived comprehensiveness and attention to special circumstances. In the reasoning model, providing reasoning decreased perceived mistake and increased perceived comprehensiveness and objectivity, but did not significantly affect attention to special circumstances. Across both models, perceived mistake negatively predicted reasonableness ratings, whereas perceived comprehensiveness, objectivity, and attention to special circumstances positively predicted reasonableness ratings.

## 6 DISCUSSION

In general, there is no evidence that attribution of advice to an AI-system adversely colored how participants judged the reasonableness of a legal opinion. At the same time, the data suggests that there are cancelling effects. Attribution of legal advice to AI seemed to reduce reasonableness ratings by influencing perceptions of attention to special circumstances, comprehensiveness and perhaps mistake. It also seemed to increase reasonableness ratings by influencing perceptions of objectivity.

These patterns are corroborated by participants' justifications for their reasonableness ratings. First, a number of participants brought up the apparent objectivity of the AI system, noting its use of "intelligent algorithms," "massive calculations of statistics," and "big data analysis". One participant emphasized the capacity of machines to "judge the entire matter without any personal emotions." This sentiment was shared by another respondent who wrote that "[the AI system] was able to correctly and rationally view the case without individuals' emotional coloring," which "allowed the case to be handled more fairly and reasonably." Others also asserted immunity to "subjective factors" to be an advantage of AI systems in the provision of legal advice.

However, objectivity was also characterized by some participants as a limitation of AI systems. One participant argued that the emotional detachment of machines compromised comprehensiveness, noting that "in real life, many things involve human feelings" and that "[the computer] could not accurately analyze all aspects of this matter." Others contended that the AI system failed to "combine with actual circumstances," to "consider realistic situations," or analyze based on "the event itself." A participant assigned to the estate division scenario thought the legal advice given to be unreasonable because "AI only considered the legal provisions granting inheritance rights to children born within and outside marriage, but did not take into account how much of the estate in Xu's family was derived from Xu himself." By comparison, several participants who received the same legal advice from a human lawyer praised her opinion as "targeted" and "comprehensive."

The quantitative and qualitative results raise the possibility of an empirical opposition between objectivity and comprehensiveness in the legal domain. While perceived objectivity may convince users to accept automated legal recommendations, it can also arouse suspicions about whether the machine has truly taken all the factual circumstances into account. As discussed earlier, law is by its very nature not a set of rigid formulas. Rules and principles do not by themselves resolve all disputes and the open texture of law means that the application of legal norms often calls for social and moral judgment. A purely objective approach, based on "intelligent algorithms" and "big data analysis" may struggle to capture the unique, messy details of human experience—the grasp of "special circumstances" and "realistic situations" that some participants found missing from legal advice attributed to an AI-system. This tension should be considered when designing and implementing AI legal advisors.

More generally, this article affirms algorithm aversion as a dynamic rather than static phenomenon. Identifying the causal pathways that shape our attitudes towards automation is important, even when those channels partially or fully offset one another. This is because changes in technological capabilities or public narratives can alter the magnitude of these pathways and thereby impact how we engage and interact with intelligent machines. Moreover, how we respond to artificial advice sometimes depends on the standards or values governing the relevant domain of human activity. Our expectations of legal counsel, for example, differ in degree if not kind from our expectations of recommendations for movies or jokes. These normative understandings may evolve over time, shifting our disposition towards algorithmic guidance on a particular problem or topic.

## 7 LIMITATIONS AND FUTURE RESEARCH

There are several limitations of our study that point toward valuable avenues for future research. First, our experiment was based on a pre-tested set of socially controversial cases drawn from Chinese civil law and was fielded in the PRC mainland. Comparative studies could explore how cultural dimensions, such as individualism-collectivism (Barnes, Zhang, & Valenzuela, 2024) or trust in institutions (Castelo, 2024), moderate attitudes towards AI-generated legal advice.

Second, the experiment presented here did not manipulate the presentation or embodiment of the AI system to users. The automated advisor was described generically as an "AI legal service platform." Extension could investigate whether anthropomorphic design elements (e.g., a conversational avatar, a name, or a more transparent explanation of the underlying model) influence perceptions of mistake, comprehensiveness, objectivity, and attention to special circumstances.

Third, our experiment measures the immediate response of ordinary people to vignettes about automated legal advisors and does not address long-term user behavior or the impact of repeated interactions on user attitudes. In addition, the outcome variable is self-reported and may be subject to cognitive biases such as social desirability (Almog, 2025). Though logistically difficult and costly to execute, longitudinal field studies can uncover how human trust and reliance on AI legal advisors evolve with continued use and experience.

Fourth, the candidate mediators were measured with single items. These items capture theoretically relevant perceptions, but they cannot capture all dimensions of the broader constructs to which those perceptions are connected. Attention to special circumstances, for instance, is closely connected to uniqueness neglect, or the concern that AI systems may fail to account for the distinctive features of individual cases (Longoni, Bonezzi, & Morewedge, 2019), as well as to the negative machine heuristic that portrays AI systems as rigid and contextually insensitive (Sundar, 2020). Comprehensiveness, the extent to which an advisor is seen to have considered all relevant elements of a case, may also overlap with notions such as diligence and analytical depth. Because these mediator measures were collected after participants saw the source label, they may reflect both participants' evaluations of the specific advice or advisor in context and broader attitudes toward AI systems that the source label brings to mind (Molina & Sundar, 2022). Validated multi-item instruments would permit a more rigorous psychometric assessment of these constructs.

Fifth, our mediation analysis suggests that the positive effect of providing reasoning on judgments of reasonableness is driven in significant part by enhanced perceptions of objectivity. This finding provokes a broader question: is the statistical relationship between reasoning and objectivity a general phenomenon or one specific to the legal domain? The importance of objectivity in the application studied here could be explained by the intimate association between neutrality and impartiality on the one hand and law on the other. In fields such as creative design or medical diagnosis, reasoning might operate to enhance perceptions of originality or empathy instead of, or in addition to, objectivity. There are therefore grounds to question whether reasoning activates the same psychological mechanisms in all spheres.

## 8 CONCLUSION

An experiment conducted on 3,348 residents in the mainland PRC finds no evidence of ordinary people exhibiting bias against AI-generated legal advice, even when that advice is socially controversial but legally correct. The source of the advice—whether attributed to an AI system or a human lawyer—had no significant effect on participants' perceptions of its reasonableness. There is, however, some indication of a more intricate underlying mechanism: attribution to an AI simultaneously enhanced perceptions of objectivity, increasing perceived reasonableness, and diminished perceptions of comprehensiveness and attention to special circumstances, decreasing perceived reasonableness. A qualitative examination of open-ended responses by participants also hints at a tension between objectivity and comprehensiveness in the legal field: the perceived objectivity that makes AI-generated advice convincing may also give rise to skepticism about its ability to grasp the fluid and circumstantial nature of human disputes. Importantly, the provision of reasoning significantly enhanced the perceived reasonableness of advice, principally, it seems, by altering perceptions of objectivity. This result implies that the explanation behind a legal opinion may be more critical to user acceptance than the human or algorithmic origin of the advice. Advancing these insights permits us to better design user interfaces and thereby optimize the benefits of technology not only for access to justice but also in other normatively salient domains.

## AUTHOR NOTE

All data generated or analyzed during this study are provided in the tables and figures and have not been previously published or used.

## REFERENCES

Almog, D. (2025, 4 26). *AI recommendations and non-instrumental image concerns.* Retrieved from arXiv: https://doi.org/10.48550/arXiv.2504.19047

Arlinghaus, C. S., Straßmann, C., & Dix, A. (2025). Increased morality through social communication or decision situation worsens the acceptance of robo-advisors. *Computers in Human Behavior: Artificial Humans*.

Barnes, A. J., Zhang, Y., & Valenzuela, A. (2024). AI and culture: Culturally dependent responses to AI systems. *Current Opinion in Psychology*.

Bohlen, L., Kruschel, S., Rosenberger, J., Zschech, P., & Kraus, M. (2025, 8 5). *Overcoming algorithm aversion with transparency: Can transparent predictions change user behavior?* Retrieved from arXiv: https://arxiv.org/abs/2508.03168

Castelo, N. (2024). Perceived corruption reduces algorithm aversion. *Journal of Consumer Psychology*.

Castelo, N., Bos, M. W., & Lehmann, D. R. (2019). Task-dependent algorithm aversion. *Journal of Marketing Research*, 809-825.

Chen, B. M., & Li, Z. (2020). How Will Technology Change The Face of Chinese Justice? *Columbia Journal of Asian Law*, 1-58.

Cohen, S. (1972). *Folk Devils and Moral Panics: The Creation of the Mods and Rockers.* London and New York: MacGibbon and Kee Ltd.

Critcher, C. (2008). Moral Panic Analysis: Past, Present and Future. *Sociology Compass*, 1127-1144.

Cullity, G. (2008). Public Goods and Fairness. *Australasian Journal of Philosophy*, 1-21.

Daschner, S., & Obermaier, R. (2022). Algorithm aversion? On the influence of advice accuracy on trust in algorithmic advice. *Journal of Decision Systems*, 77-97.

de Jong, S., Bos, M., van Berkel, N., & Lamers, M. (2025). Algorithm appreciation or aversion: The effects of accuracy disclosure on users' reliance on algorithmic suggestions. *Behaviour & Information Technology*, 1-20.

Dietvorst, B. J., Simmons, J. P., & Massey, C. (2015). Algorithm aversion: People erroneously avoid algorithms after seeing them err. *Journal of Experimental Psychology: General*, 114-126.

Dietvorst, B. J., Simmons, J. P., & Massey, C. (2018). Overcoming algorithm aversion: People will use imperfect algorithms if they can (even slightly) modify them. *Management Science*, 1155-1170.

Dworkin, R. (1986). *Law's Empire.* Harvard University Press.

Flanagan, B., Almeida, G., Chen, D., & Gitahi, A. (2025). The rule of law or the rule of robots? Nationally representative survey evidence from Kenya. *Information & Communications Technology Law*.

Goode, E., & Ben-Yehuda, N. (1994). *Moral Panics: The Social Construction of Deviance.* Wiley-Blackwell: Oxford, England.

Hagan, M. (2024). Towards human-centered standards for legal help AI. *Philosophical Transactions of the Royal Society A: Mathematical, Physical and Engineering Sciences*.

Harasta, J., Novotná, T., & Savelka, J. (2024). It cannot be right if it was written by AI. *Artificial Intelligence and Law*.

Hart, H. L. (1958). Positivism and the separation of law and morals. *Harvard Law Review*, 593-629.

Hart, H. L. (1961). *The Concept of Law.* Oxford University Press.

Hayes, A. F. (2013). *Introduction to mediation, moderation, and conditional process analysis: A regression-based approach.* New York: The Guilford Press.

Heimstad, S. B., Wien, A. H., & Gaustad, T. (2025). Machine heuristic in algorithm aversion: Perceived creativity and effort of output created by or with artificial intelligence. *Computers in Human Behavior: Artificial Humans*.

Hou, Y. T.-Y., & Jung, M. F. (2021). Who is the expert? Reconciling algorithm aversion and algorithm appreciation in AI-supported decision making. *Proceedings of the ACM on Human-Computer Interaction* (pp. 1-25). ACM Journals.

Hovland, C. I., Janis, I. L., & Kelley, H. H. (1953). *Communication and persuasion; psychological studies of opinion change.* New Haven: Yale University Press.

Imai, K., Keele, L., & Yamamoto, T. (2010). Identification, inference and sensitivity analysis for causal mediation effects. *Statistical Science*, 51-71.

Imbens, G. W., & Rubin, D. B. (2015). *Causal Inference for Statistics, Social, and Biomedical Sciences: An Introduction.* Cambridge University Press.

Jussupow, E., Benbasat, I., & Heinzl, A. (2020). Why are we averse towards algorithms? A comprehensive literature review on algorithm aversion. *the 28th European Conference on Information Systems (ECIS 2020)* (pp. 1-17). Marrakech, Morocco: ECIS.

Kizilcec, R. F. (2016). How Much Information?: Effects of Transparency on Trust in an Algorithmic Interface. *Proceedings of the 2016 CHI Conference on Human Factors in Computing Systems (CHI '16)* (pp. 2390-2395). New York: Association for Computing Machinery.

Kunkel, J., Donkers, T., Michael, L., Barbu, C.-M., & Ziegler, J. (2019). Let Me Explain: Impact of Personal and Impersonal Explanations on Trust in Recommender Systems. *Proceedings of the 2019 CHI Conference on Human Factors in Computing Systems (CHI '19)* (pp. 1-12). New York: Association for Computing Machinery.

Liu, Y., & Wang, T. (2025). Treating differently or equally: A study exploring attitudes towards AI moral advisors. *Technology in Society*.

Logg, J. M., Minson, J. A., & Moore, D. A. (2019). Algorithm appreciation: People prefer algorithmic to human judgment. *Organizational Behavior and Human Decision Processes*, 90–103.

Longoni, C., Bonezzi, A., & Morewedge, C. K. (2019). Resistance to medical artificial intelligence. *Journal of Consumer Research*, 629-650.

Lu, J. (2016). On randomization-based and regression-based inferences for 2K factorial designs. *Statistics & Probability Letters*, 72-78.

Magni, F., Park, J., & Chao, M. M. (2024). Humans as creativity gatekeepers: Are we biased against AI creativity? *Journal of Business and Psychology*, 643-656.

Mayer, R. C., Davis, J. H., & Schoorman, F. D. (1995). An Integrative Model of Organizational Trust. *The Academy of Management Review*, 709-34.

Molina, M. D., & Sundar, S. S. (2022). When AI moderates online content: effects of human collaboration and interactive transparency on user trust. *Journal of Computer-Mediated Communication*, 1-12.

Mosier, K. L., Skitka, L. J., Heers, S., & Burdick, M. (1998). Automation bias: Decision making and performance in high-tech cockpits. *International Journal of Aviation Psychology*, 47-63.

Ning, X., Lu, Y., Li, W., & Gupta, S. (2024). How transparency affects algorithmic advice utilization: The mediating roles of trusting beliefs. *Decision Support Systems*, 1-15.

Parasuraman, R., & Riley, V. (1997). Humans and automation: Use, misuse, disuse, abuse. *Human Factors*, 230-253.

Prahl, A., & Van Swol, L. M. (2017). Understanding algorithm aversion: When is advice from automation discounted? *Journal of Forecasting*, 691-702.

Qin, X., Zhou, X., Chen, C., Wu, D., Zhou, H., Dong, X., . . . Lu, J. G. (2025). AI Aversion or Appreciation? A Capability – Personalization Framework and a Meta-Analytic Review. *Psychological Bulletin*, 580-599.

Raz, J. (1979). *The Authority of Law: Essays on Law and Morality.* Oxford University Press.

Riess, H., & Kraft-Todd, G. (2014). E.M.P.A.T.H.Y.: A tool to enhance nonverbal communication between clinicians and their patients. *Academic Medicine*, 1108-1112.

Rosseel, Y. (2012). lavaan: An R Package for Structural Equation Modeling. *Journal of Statistical Software*, 1-36.

Schneiders, E., Seabrooke, T., Krook, J., Hyde, R., Leesakul, N., Clos, J., & Fischer, J. (2025). Objection overruled! Lay people can distinguish large language models from lawyers, but still favour advice from an LLM. *2025 CHI Conference on Human Factors in Computing Systems (CHI '25).* Yokohama, Japan: ACM.

Seabrooke, T., Schneiders, E., Dowthwaite, L., Krook, J., Leesakul, N., Clos, J., . . . Fischer, J. (2024). A survey of lay people's willingness to generate legal advice using large language models (LLMs). *Second International Symposium on Trustworthy Autonomous Systems (TAS '24)* (pp. 1-5). Austin, TX: ACM.

Shin, D. (2021). The effects of explainability and causability on perception, trust, and acceptance: Implications for explainable AI. *International Journal of Human-Computer Studies*, 1-10.

Spencer, S. J., Zanna, M. P., & Fong, G. T. (2005). Establishing a causal chain: why experiments are often more effective than mediational analyses in examining psychological processes. *Journal of Personality and Social Psychology*, 845-51.

Sundar, S. (2020). Rise of Machine Agency: A Framework for Studying the Psychology of Human–AI Interaction (HAII). *Journal of Computer-Mediated Communication*, 74–88.

Turel, O., & Kalhan, S. (2023). Prejudiced against the machine? Implicit associations and the transience of algorithm aversion. *MIS Quarterly*, 1369-1394.

VanderWeele, T. (2015). *Explanation in Causal Inference: Methods for Mediation and Interaction .* Oxford: Oxford University Press.

Waismann, F. (1945). Verifiability. *Proceedings of the Aristotelian Society*, (pp. 119-150).

Yalcin, G., Lim, S., Puntoni, S., & van Osselaer, S. M. (2022). Thumbs Up or Down: Consumer Reactions to Decisions by Algorithms Versus Humans. *Journal of Marketing Research*, 696-717.